\documentstyle[prl,aps,epsfig]{revtex}
\begin{document} 

\def\slash#1{\setbox0=\hbox{$#1$}#1\hskip-\wd0\dimen0=5pt\advance
\dimen0 by-\ht0\advance\dimen0 by\dp0\lower0.5\dimen0\hbox
to\wd0{\hss\sl/\/\hss}}
\def \gam {\frac{ N_f N_cg^2_{\pi q\bar q}}{8\pi} } 
\def \gamm {N_f N_cg^2_{\pi q\bar q}/(8\pi) } 
\def \be {\begin{equation}} 
\def \ba {\begin{eqnarray}} 
\def \ee {\end{equation}} 
\def \ea {\end{eqnarray}} 
\def\ct {{C_{10}}}
\def\cne {C_9^{\rm eff}}
\def\cse {C_7^{\rm eff}}
\def \sh {{\hat{q}}^2}
\def \uh {\varsigma} 
\def\mvh {{\hat{m}}_\phi}
\def\mlh {{\hat{m}}_\ell}
\def\mbh {{\hat{m}}_b}
\def\msh {{\hat{m}}_s}
\def \im {{\rm Im}} 
\def \re {{\rm Re}} 
\def \Tr {{\rm Tr}} 
\def \P {$0^{-+}$} 
\def \S {$0^{++}$} 
\def\d{{\rm d}}

\begin{titlepage}
\null
\vspace{2cm}
\begin{center}
\Large\bf 
The exclusive $B_s\to\phi\mu^+\mu^-$ process in a constituent  
quark model
\end{center}
\vspace{1.5cm}

\begin{center}
\begin{large}
Aldo Deandrea\\
\end{large}
\vspace{0.5cm}
Institut de Physique Nucl\'eaire, Universit\'e de Lyon I\\ 
4 rue E.~Fermi, F-69622 Villeurbanne Cedex, France \\
\vspace{0.7cm}
\begin{large}
Antonio D. Polosa\\
\end{large}
\vspace{0.5cm}
Physics Department, University of Helsinki, \\
POB 64, FIN--00014, Finland
\end{center}

\vspace{1.3cm}

\begin{center}
\begin{large}
{\bf Abstract}\\[0.5cm]
\end{large}
\parbox{14cm}{We consider the exclusive $B_s\to\phi\mu^+\mu^-$ process in the 
standard model using a constituent  quark loop model approach together with a simple
parameterization of the quark dynamics. The model allows to compute the decay
form factors and therefore can give predictions for the decay rates, the
invariant mass spectra and the asymmetries. This process is suppressed in
the standard model but can be enhanced if new physics beyond the standard model
is present, such as flavor-violating supersymmetric models. It constitutes
therefore an interesting precision test of the standard model at forthcoming
experiments.}
\end{center}

\vspace{1.5cm}
\vfil
\noindent
HIP-2001-16/TH\\
LYCEN-2001-27\\
May 2001
\end{titlepage}

\title{The exclusive $B_s\to\phi\mu^+\mu^-$ process in a constituent  
quark model} 
\author{A. Deandrea$^{(a)}$ and A.D. Polosa$^{(b)}$} 
\address{(a) Institut de Physique Nucl\'eaire, Universit\'e Lyon I, 
4 rue E.~Fermi, F-69622 Villeurbanne Cedex, France\\ 
(b) Physics Department, University of Helsinki, POB 64, FIN--00014,  
Finland} 
\date{May, 2001} \maketitle 
\begin{abstract} 
We consider the exclusive $B_s\to\phi\mu^+\mu^-$ process in the standard 
model using a constituent  quark loop model approach together with a simple
parameterization of the quark dynamics. The model allows to compute the decay
form factors and therefore can give predictions for the decay rates, the
invariant mass spectra and the asymmetries. This process is suppressed in
the standard model but can be enhanced if new physics beyond the standard model
is present, such as flavor-violating supersymmetric models. It constitutes
therefore an interesting precision test of the standard model at forthcoming
experiments.\\ 
\vskip 0.05cm 
\noindent Pacs numbers: 13.25.Hw, 12.39.Hg, 14.40.Nd\\ \noindent 
HIP-2001-16/TH , LYCEN-2001-27\vskip 0.90cm 
\end{abstract} 

\section{Introduction} 
We discuss the $B_s\to\phi\mu^+\mu^-$ exclusive 
process using a Constituent Quark--Meson model (CQM) \cite{nc} based on
quark--meson interactions. Quark--meson interaction vertices
can be obtained by partial bosonization of a Nambu--Jona-Lasinio type
four--quark interaction vertices for heavy and light quarks \cite{qmi}. 
In this model the transition amplitudes are
evaluated by computing diagrams in which heavy and light mesons
are attached to quark loops. Moreover, the light chiral symmetry 
restrictions and the heavy quark spin-flavour symmetry dictated by the
Heavy-Quark-Effective-Theory (HQET) are both implemented. 

Flavor Changing Neutral Current processes (FCNC), like $B_s\to\phi\mu^+\mu^-$
where the $b$ quark is transformed into a $s$ quark by a neutral current, are
absent in  the Standard Model (SM) at the tree level. This makes the effective
strengths of such processes small. In addition, these transitions are
dependent on the weak mixing angles of the Cabibbo-Kobayashi-Maskawa (CKM)
matrix and can be suppressed also due to their proportionality  
to small CKM elements.   
If the quark masses in the FCNC loop diagrams are close to each 
other, the GIM mechanism is effective and this implies that FCNC 
charm decay transitions are very suppressed. On the other hand, 
FCNC beauty decays should be observable at LHC. These decays, 
known as rare $B$ decays, are extremely interesting for the study 
of new physics. Indeed, any observed enhancement of their 
branching ratio could be the trace of some no-SM mechanism. 
Besides the sensitivity of 
rare $B$ decays to new physics, their study is very important  
in the context of the SM for a comparative determination of the 
CKM matrix elements $V_{tb},V_{td}$ and $V_{ts}$\cite{sanda}: 
these quantities can be measured directly in top quark decays. 
 
Here we study, in the CQM approach, the 
exclusive process $B_s\to\phi\mu^+\mu^-$. Our aim is to provide an 
independent determination of this process in a different context 
with respect to that of QCD Sum Rules, where it has been studied 
first \cite{ball}. Exclusive semileptonic decays are in general 
more complicated than inclusive modes on the theoretical ground 
since they require the determination of form factors. On the other 
hand they are very promising on the experimental side, being 
accessible to a large number of ongoing and future experiments. 
The method we apply can be used for the calculation 
of other processes like $B_s\to K^*\ell^+\ell^-$ or $B_{u,d}\to 
K^*\ell^+\ell^-$, $B_{d,s}\to \eta\ell^+\ell^-$ following the same 
techniques described in this paper. 
 
The CQM model has turned out to be particularly suitable for the 
study of heavy meson decays. Its Lagrangian describes 
the vertices (heavy meson)-(heavy quark)-(light 
quark) \cite{nc}, and transition amplitudes are computable via simple 
constituent quark loop diagrams in which mesons appear as external 
legs. In the case of the process at hand, 
two interfering diagrams contribute to the transition amplitude, 
see Fig. 1 and Fig. 2. They correspond to the two
physically  distinct alternatives in which the FCNC effective vertex is either 
attached directly to the quark loop or via an intermediate heavy 
meson state (the matrix elements factorize into an hadronic and 
leptonic part). We will compute the two 
contributions to the process and discuss their relative weight. 
 
\section{Effective Hamiltonian and form factors}
At the quark level, the rare semileptonic decay $b\to 
s\ell^+\ell^-$ can be described in terms of the effective 
Hamiltonian obtained by integrating out the top quark and 
$W^{\pm}$ bosons: 
\begin{equation} 
{\cal H}_{\rm 
eff}=-4\frac{G_F}{\sqrt{2}}V_{ts}^*V_{tb}\sum_{i=1}^{10}C_i(\mu){\cal 
O}_i(\mu). 
\label{hami}
\end{equation} 
We will use the Wilson coefficients calculated in the naive 
dimensional regularization scheme\cite{buras} (see Table I). 
The numerical values used in the calculations are given in Table
II. The Hamiltonian (\ref{hami}) leads to the following free
quark decay  amplitude\cite{ball}: 
\begin{eqnarray} {\cal M}(b\to 
s\ell^+\ell^-)&=&\frac{G_F \alpha}{\sqrt{2}\pi} V_{ts}^* 
V_{tb}\left[\frac{C_9^{\rm 
eff}}{2}(\bar{s}\gamma_\mu(1-\gamma_5)b)(\bar{\ell}\gamma^\mu\ell)+ 
\frac{C_{10}}{2}(\bar{s}\gamma_\mu(1-\gamma_5)b)(\bar{\ell}\gamma^\mu
\gamma_5\ell)\nonumber\right.\\ 
&-& \left.\frac{m_b}{q^2}C_7^{\rm 
eff}(\bar{s}i\sigma_{\mu\nu}q^\nu(1+\gamma_5)b)(\bar{\ell}\gamma^\mu\ell)
\right], 
\label{def} 
\end{eqnarray} 
where $C_i(\mu=m_b)$ are the Wilson coefficients  which act as 
effective coupling constants in the 4-Fermi formulation of the 
interactions. Here $C_7^{\rm eff}=C_7-C_5/3-C_6$. 
Short-distance Wilson coefficients are redefined in 
such a way to incorporate certain long-distance effects from the 
matrix elements of four-quark operators ${\cal O}_i$ with $1\leq 
i\leq 6$. $C_9^{\rm eff}$, the Wilson coefficient of ${\cal 
O}_9=e^2/16\pi^2 (\bar{s}\gamma_\mu Lb)(\bar{\ell}\gamma^\mu 
\ell)$, is defined in terms of these matrix elements in 
the Appendix. In order to compute the $\langle \phi 
\ell^+\ell^-|{\cal M}(b\to s\ell^+\ell^-)| B_s\rangle$, we need 
the following form factors parameterization for the $V-A$ terms: 
\begin{eqnarray} 
\label{eq:monsterm} <\phi(\epsilon,p)|\bar{s}\gamma_\mu 
(1-\gamma_5)b|\bar{B_s}(p_B)> & = & \frac{2 
V(q^2)}{m_B+m_{\phi}}\epsilon_{\mu\nu\alpha\beta} 
\epsilon^{*\nu}p_B^\alpha p^{\beta}\nonumber\\ &-& i 
\epsilon^{*}_{\mu}(m_{B} + m_{\phi})  A_{1}(q^{2}) \nonumber\\ &+& 
i (\epsilon^{*}\cdot q) \frac{(p_B + p)_{\mu}}{m_B + m_{\phi}} 
A_{2}(q^{2}) \label{ffsemilep} \\ 
&+& i  (\epsilon^{*}\cdot  q) \frac{2  m_{\phi}}{q^{2}} q_{\mu} 
[A_{3}(q^{2})  - A_{0}(q^{2})], \nonumber 
\end{eqnarray} 
where $q=p_B-p$ and: 
\begin{equation} 
A_{3}(q^{2})  = \frac{m_{B} + m_{\phi}}{2  m_{\phi}} A_{1}(q^{2}) 
- \frac{m_{B} - m_{\phi}}{2  m_{\phi}} A_{2}(q^{2}), 
\end{equation} 
with the condition: 
\begin{equation} 
A_3(0)=A_0(0). \label{condi} 
\end{equation} 
The form factors needed for the magnetic term in (\ref{def}) are 
defined by: 
\begin{eqnarray} 
<\phi(\epsilon,p)|\bar{s}\sigma_{\mu\nu} 
q^{\nu}(1+\gamma_5)b|\bar{B_s}(p_B)> & = & 
i\epsilon_{\mu\nu\alpha\beta}\epsilon^\nu p_B^{\alpha}p^{\beta} 
2T_1(q^2)\nonumber\\ &+& 
T_2(q^2)\left[\epsilon_\mu(m_B^2-m_\phi^2)-(\epsilon\cdot 
p_B)(p_B+p)_\mu\right]\nonumber\\ &+& T_3(q^2)\left[(\epsilon\cdot 
p_B)q_\mu-\frac{q^2}{(m_B^2-m_\phi^2)}(p_B+p)_\mu\right], 
\label{param} 
\end{eqnarray} 
with: 
\begin{equation} 
T_1(0)=T_2(0). 
\end{equation} 
These seven form factors can be 
calculated with the aid of a Constituent-Quark-Meson (CQM) model 
in which the interactions 
$B_s-\phi-(V,A,T)$current are modeled by effective constituent 
quark loop diagrams. The $\phi$ is considered to have a 
photon-like interaction with the light $s$ degrees of freedom 
described by: 
\begin{equation} 
{\cal L}=ig_{\phi ss}\bar{s}\gamma_\mu s\phi^\mu,
\label{otto} 
\end{equation} 
where $g_{\phi ss}= g_\phi/3$, the factor of $3$ coming from the 
charge of the $s$ quark. $g_\phi$ is extracted from the measured 
electronic width of the $\phi$ via $\Gamma(\phi\to 
e^+e^-)=(4\pi\alpha^2)/3 (1/g_\phi^2)m_\phi$. 
This way of modeling the interaction derives from the Vector Meson 
Dominance (VMD). Vector meson dominance can be obtained from the 
effective four--quark theory of the Nambu--Jona-Lasinio type for light quarks
once electro--weak interactions are added. One can show \cite{Ebert:1983pk}
that the coupling of the electromagnetic field $A_\mu$ with quarks is turned
into a direct coupling of photons and neutral vector mesons in the effective
theory, and this reproduces the VMD term. Eq. (\ref{otto}) can therefore be
interpreted as a quark--quark--light-meson vertex of this kind.  

As stated above, there are two kind of contributions to the form  factors,
depicted respectively in Fig. 1 and Fig. 2. In the first  one the current is
directly attached to the loop of quarks. In the  second, there is a
intermediate state between the current and the  $B_s\phi$ system. This
intermediate state is a $1^-$ or $1^+$  heavy meson with a $s$ valence quark.
The Feynman rules for  computing these diagrams have been discussed in
\cite{nc} and the  extension of the model to the strange quark sector has been 
developed in \cite{ultimo}. Consider for example the {\it direct} 
diagram in Fig. 1. The model allows to extract the direct 
contributions to the form factors $V^D,A_1^D,..,T_1^D,..$ through 
the calculation of the loop integral: 
\begin{equation} 
\sqrt{Z_H m_B} \frac{iN_c}{16\pi^4}\int^{\rm reg} d^4\ell 
\frac{{\rm Tr}\left[ (\gamma\cdot\ell +m)(g_{\phi ss} 
\gamma\cdot\epsilon)(\gamma\cdot(\ell+p)+m)({\rm V,A,T}) 
\frac{1+\gamma\cdot v}{2}\gamma_5 
\right]}{(\ell^2-m^2)((\ell+p)^2-m^2)(v\cdot\ell+\Delta_H)}, 
\label{integral} 
\end{equation} 
where $V,A,T$ denote respectively Vector, Axial-Vector and Tensor 
($T=\sigma_{\mu\nu}(1+\gamma_5)$) currents, $\ell$ is the momentum 
running in the loop, $p=m_\phi v^\prime$ is the momentum of the 
$\phi$, $m$ the constituent quark mass of the strange quark (we 
have $m=510$ MeV), $v$ is the four velocity of the incoming heavy 
$(0^-)$ meson; the heavy quark propagator and the heavy meson 
field expressions from Heavy Quark Effective Theory (HQET) are 
taken into account. The constant $\Delta_H$ is defined as the 
$m_B-m_b$ difference (between the mass of the heavy meson and the 
constituent heavy quark mass) and represents, together with the 
cutoffs, the basic input parameter of the model. Following 
\cite{ultimo}, here we assume $\Delta_H=0.6$ GeV. $Z_H$ is  
the coupling constant of the heavy meson field $H$ (using the 
notations of HQET where $H$ represents the $(0^-,1^-)$ heavy meson 
multiplet) with the quark propagators. Integrals like 
(\ref{integral}) are obviously divergent. We define the model with 
the proper time regularization procedure which consists in 
exponentiating the light quark propagators in the Euclidean space, 
and introducing ultraviolet ($\Lambda$) and infrared ($\mu$) 
cutoffs in the proper time variable. In our numerical analysis 
$\Lambda=1.25$ GeV and $\mu=0.51$ GeV. The results are quite 
stable against $10-15\%$ variations of the cutoff values. We again 
refer to \cite{nc,ultimo} for a discussion on the determination 
and the physical meaning of these parameters. 

\subsection{Direct form factors} 
The CQM expressions for the contributions to the form factors, 
derived by the {\it direct} diagram calculations with $V$ and $A$ 
currents (see Fig. 1), are the following: 
\begin{eqnarray} 
V^{D}(q^{2}) &=& -g_{\phi ss} \sqrt{\frac{Z_H} {m_{B_s}}} \left( 
\Omega_1 - m Z \right) (m_{B_s} + m_{\phi})\\  
A^{D}_1 (q^{2}) &=& 
2g_{\phi ss}\sqrt{Z_H m_{B_s}} \frac{1}{m_{B_s} + m_{\phi}} 
\nonumber 
\\ && \left[ (m^2 + m m_{\phi}\omega) Z -{\omega} 
m_{\phi}\Omega_1 - m_\phi \Omega_2 -2 \Omega_3 -\nonumber 
\right.\\ && \left. \Omega_4 -\Omega_5 -2 {\omega} \Omega_6 
\right]\\ 
A^{D}_2(q^{2}) &=& g_{\phi ss}\sqrt{\frac{Z_H}{m_{B_s}}} 
\left( m Z -\Omega_1 - 2 \frac{\Omega_6}{m_\phi} \right) (m_{B_s} 
+ m_{\phi})\\ 
A^{D}_0 (q^{2}) &=& -\frac{g_{\phi ss}}{m_\phi} \sqrt{Z_H m_{B_s}} 
\left[\Omega_1 \left( m_\phi {\omega} +2 m 
\frac{q^2}{m_{B_s}^2} -\frac{r_1}{m_{B_s}}\right) +  m_\phi 
\Omega_2 + \nonumber \right.\\ &&\left. 2\Omega_3 + \Omega_4 
\left(1- 2 \frac{q^2}{m_{B_s}^2}\right) + \Omega_5 + 2\Omega_6 
\left( \omega- \frac {r_1}{m_{B_s} m_\phi} 
\right)-\nonumber\right.\\ &&\left.  Z \left(m^2  - m 
\frac{r_1}{m_{B_s}}  + m m_\phi {\omega}\right) \right], 
\end{eqnarray} 
where: 
\begin{equation} 
{\omega}=\frac{m_{B_s}^2+m_\phi^2-q^2}{2 m_{B_s} m_\phi}, 
\label{writt} 
\end{equation} 
and: 
\begin{equation} 
r_1=\frac{m_{B_s}^2-q^2-m^2_\phi}{2} .
\label{r2} 
\end{equation} 
The definitions of the functions $Z,\Omega_i$ are the
following:  
\begin{eqnarray} 
\label{eq:beyond} Z&=&\frac{iN_c}{16\pi^4}\int^{\rm reg} 
\frac{d^4\ell}{(\ell^2-m^2)((\ell+p)^2-m^2)(v\cdot 
\ell+\Delta_1+i\epsilon)}\\ Z^{\mu}&=&\Omega_1 v^\mu +\Omega_2 
v^{\prime \mu}\\ Z^{\mu\nu}&=&\Omega_3g^{\mu\nu} +\Omega_4 v^\mu 
v^\nu+ \Omega_5 v^{\prime\mu}v^{\prime\nu}+\Omega_6[v^\mu 
v^{\prime\nu}+v^{\prime\mu}v^\nu], 
\end{eqnarray} 
where $Z^\mu$ and $Z^{\mu\nu}$ mean that in the $Z$ integral we 
are considering $\ell^\mu$ and $\ell^\mu\ell^\nu$ numerators. $v$ 
is the four velocity of the incoming meson, $v^\prime$ the four 
velocity of the $\phi$. In the computations, $v$ and $v^\prime$ 
are related by $v\cdot v^\prime=(\Delta_1-\Delta_2)/m_\phi$ where 
$\Delta_1$ is $v\cdot k$, $k$ being the residual momentum of the 
heavy quark. The explicit expressions for the integrals $Z$, 
$\Omega_i$ are algebraic combinations of some $I_i$ integrals 
which are given in the Appendix. They are in general 
functions of the two parameters, $\Delta_1$ and $\Delta_2$. In the 
case of the direct diagrams in Fig. 1, $\Delta_2=\Delta_1-m_\phi 
\omega$, where $\omega$ has been written in 
(\ref{writt}). 

\subsection{Polar form factors}
The {\it polar} contributions to the form factors come from those 
CQM diagrams in which the weak current is coupled to $B_s$ through 
an heavy meson intermediate state, see Fig. 2. The form factor 
will then have a typical polar behavior: 
\begin{equation} 
F^P(q^2)=\frac{F^P(0)}{1~-~q^2/m_P^2}, \label{eq:fp} 
\end{equation} 
where $m_P$ is the mass of the intermediate virtual heavy meson 
state. Pole masses are given in Table IV. This behavior is
certainly valid near the pole; we will  assume that it is valid all over the
$q^2$ range that we want to explore, {\it i.e.}, also for small $q^2$ values.
We find:  \begin{eqnarray} 
V^P (0)&=&  -\sqrt{2} g_V \lambda {\hat F} \frac{m_{B_s} +m_\phi}{ 
m_{B_s}^{3/2}}\\ 
A^P_1 (0) &=&  \frac{\sqrt{2 m_{B_s}}g_V {\hat 
F}^+}{m_{B_s^{**}} (m_{B_s}+m_{\phi})} (\zeta-2\mu {\omega} 
m_{\phi})\\ 
A^P_2 (0) &=& -\sqrt{2} g_V \mu {\hat F}^+ 
\frac{\sqrt{m_{B_s}} (m_{B_s}+m_\phi)}{m_{B_s^{**}}^2}, 
\end{eqnarray} 
where ${\omega}=m_{B_s}/(2 m_\phi)$, while $g_V=m_\phi/f_\pi$ 
\cite{report}. By $B_s^{**}$ we mean the $1^+$ state of the $S$ 
multiplet $(0^+,1^+)$ of HQET. The mass of $B_s^{**}$ is taken by 
\cite{matsuki} to be $m_{B_s^{**}}=5.76$ GeV. At present there 
are no experimental information about this state. 
 
As for $A^P_0 (q^2)$, we have to impose the condition 
(\ref{condi}); a choice is: 
\begin{equation} 
A^P_0 (q^2)= A^P_3(0) + g_V \beta {\hat F} \frac{1}{m_\phi\sqrt{2 
m_{B_s}}} \frac{q^2}{m^2_{B_s}-q^2}. \label{eq:AP0} 
\end{equation} 
 
$\hat{F}$ and $\hat{F}^+$ are the lepton decay constant of the $H$ 
and $S$ HQET multiplets \cite{report}: 
\begin{eqnarray} 
\label{eq:matf1} \langle {\rm VAC}|{\bar q} \gamma^\mu \gamma_5 Q 
|H\rangle  & =&i \sqrt{m_H} r^\mu {\hat F}\\ \label{eq:matf2} 
\langle {\rm VAC}|{\bar q} \gamma^\mu  Q |S\rangle & =&i 
\sqrt{m_S} r^\mu {\hat F}^+, 
\end{eqnarray} 
where we use $r=v$ (the heavy quark velocity) for $H=0^-,S=0^+$ and
$r=\epsilon$ (the polarization vector of the heavy meson) for  $H=1^-,S=1^+$.
One finds:  
\begin{eqnarray} 
\hat{F}&=&2\sqrt{Z_H}(I_1+(\Delta_H+m)I_3(\Delta_H))\\ 
\hat{F}^+&=&2\sqrt{Z_S}(I_1+(\Delta_S-m)I_3(\Delta_S)). 
\end{eqnarray} 
The numerical table connecting $\Delta_H$ and $\Delta_S$ values 
has been discussed in \cite{ultimo}. We notice that: 
\begin{equation} 
f_{B_s}=\frac{\hat{F}}{\sqrt{m_{B_s}}}, 
\end{equation} 
and numerically, we find: 
\begin{equation} 
f_{B_{s}}= 180^{+17}_{-14}\;{\rm MeV}, 
\end{equation} 
which is in reasonable agreement with the result from QCD sum 
rules \cite{ball} and from lattice\cite{wise}, according to which 
$f_{B_s}\simeq 190$ MeV. The theoretical error in the 
determination of $f_{B_s}$ comes from varying the 
$\Delta_1=\Delta_H$ parameter in the range of values 
$\Delta_H=0.5,0.6,0.7$ GeV. 
 
The CQM explicit expressions for the strong constants 
$\lambda,\beta,\mu,\zeta$, parameterizing the strong couplings 
$HH\phi$ and $HS\phi$ according to the interaction Lagrangians 
discussed in \cite{report}, are given by: 
\begin{eqnarray} 
\lambda &=&\frac{g_{\phi ss}}{\sqrt{2} g_V} Z_H (-\Omega_1+ m Z)\\ 
\beta &=&\sqrt{2}\frac{g_{\phi ss}}{g_V}   Z_H [2 m \Omega_1+ 
m_\phi \Omega_2 + 2 \Omega_3 + \Omega_4 + \Omega_5- m^2 Z ]. 
\end{eqnarray} 
Here the functions $Z$, $\Omega_j$ are computed with 
$\Delta_1=\Delta_2=\Delta_H$, $x=m_\phi$, $\omega=m_\phi/(2m_{B_s})$ 
where one takes the first $1/m_Q$ correction to $\omega=0$. 
Moreover: 
\begin{eqnarray} 
\mu &=& \frac{g_{\phi ss}}{\sqrt{2} g_V} \sqrt{Z_H Z_S}\left( 
-\Omega_1- 2 \frac{\Omega_6}{m_\phi}+ m Z\right) \\ \zeta &=& 
\frac{\sqrt{2}g_{\phi ss}}{g_V} \sqrt{Z_H Z_S} \left(m_\phi 
\Omega_2 +2 \Omega_3 +\Omega_4 +\Omega_5 -m^2 Z \right), 
\end{eqnarray} 
where $\Delta_1=\Delta_H$, $\Delta_2=\Delta_S$, $x=m_\phi$  and 
$\omega=(\Delta_1-\Delta_2)/{m_\phi}$. The suffix $S$ indicates 
the $(0^+,1^+)$ multiplet of HQET. 

\subsection{Direct tensor form factors} 
Let us now turn to the $T_i$ form factors. The contributions to 
$T_i$ coming from the direct diagrams in Fig. 1 are labeled by 
$T_i^D$. A calculation of the loop integral 
(\ref{integral}), in which we retain the $T$ current of the 
electromagnetic penguin operator, allows to extract the $T_i^D$ by 
comparison with (\ref{param}). We obtain the following results: 
\begin{eqnarray} 
T_1^D(q^2)&=&-g_{\phi ss}\sqrt{Z_H m_{B_s}} 
\left[\Omega_1+\frac{\Omega_2 
m_\phi}{m_{B_s}}-Z m+\frac{1}{m_{B_s}}\left(2 
\Omega_3+\Omega_4\nonumber\right.\right.\nonumber\\ 
&+&\left.\left. 
\Omega_5+\frac{2P}{m_{B_s}m_\phi}\Omega_6\right)-\frac{m^2Z}{m_{B_s}} 
\right]\\ 
T_2^D(q^2)&=&-\frac{2 g_{\phi 
ss}}{(m_{B_s}^2-m_\phi^2)}\sqrt{Z_H 
m_{B_s}}\left[K\Omega_1+\frac{m_\phi\Omega_2 J}{m_{B_s}}- K Z m 
+\frac{J}{m_{B_s}}\left(2\Omega_3+\Omega_4\right.\right.\nonumber\\ 
&+&\left.\left.\Omega_5+\frac{2P}{m_{B_s}m_\phi}\Omega_6\right)-\frac{m^2 
Z J}{m_{B_s}} \right]\\ 
T_3^D(q^2)&=&-g_{\phi ss}\sqrt{Z_H 
m_{B_s}}\left[\Omega_1+\frac{2\Omega_6}{m_{B_s}^2 
m_\phi}(J+K-P)-\frac{m_\phi}{m_{B_s}}\Omega_2 - Z m 
+\frac{m^2Z}{m_{B_s}}\right.\nonumber\\ &-&\left. 
\frac{1}{m_{B_s}}\left(2\Omega_3+\Omega_4+\Omega_5 \right)\right], 
\end{eqnarray} 
where: 
\begin{eqnarray} 
J&=&\frac{m_{B_s}^2-m_\phi^2+q^2}{2}\\ 
K&=&\frac{m_{B_s}^2-m_\phi^2-q^2}{2}\\ 
P&=&\frac{m_{B_s}^2+m_\phi^2-q^2}{2}, 
\end{eqnarray} 
and the following consistency condition is satisfied: 
\begin{equation} 
T_1^D(q^2=0)=T_2^D(q^2=0). 
\end{equation} 

\subsection{Polar tensor form factors}
The calculation of polar contributions to the $T_i$ form factors 
follows the computation in \cite{report}. In the latter reference 
a different parameterization of the tensor current 
matrix element is used (and also a different definition of 
$\sigma_{\mu\nu}$ from the one adopted in this paper; namely the 
$\sigma_{\mu\nu}$ is defined without an overall $i$): 
\begin{eqnarray} 
\langle 
\phi(p,\epsilon)|\bar{s}\sigma_{\mu\nu}(1+\gamma_5)b|\overline{B}_s(p_B)
\rangle &=&i[(p_{B\mu} \epsilon_\nu-p_{B\nu} 
\epsilon_\mu-i\epsilon_{\mu\nu\lambda\sigma}p_B^\lambda\epsilon^\sigma)
A(q^2)\nonumber\\  &+&(p_{\mu} \epsilon_\nu-p_{\nu} 
\epsilon_\mu-i\epsilon_{\mu\nu\lambda\sigma}p^\lambda\epsilon^\sigma)
B(q^2)\nonumber\\  &+&(\epsilon\cdot 
p_B)(p_{B\mu}p_{\nu}-p_{B\nu}p_{\mu}-i\epsilon_{\mu\nu\lambda\sigma}
p_B^\lambda p^\sigma)H(q^2)]. 
\end{eqnarray} 
 
The relations between the form factors $A(q^2),B(q^2),H(q^2)$ and 
our form factors $T_1(q^2),T_2(q^2)$ and $T_3(q^2)$ are given by: 
\begin{eqnarray} 
T_1(q^2)&=&-\frac{i}{2}\left[A(q^2)+B(q^2)\right] \\ 
T_2(q^2)&=&-\frac{i}{2}\left[A(q^2)+B(q^2)\right] 
-\frac{i}{2}\left[A(q^2)-B(q^2)\right] \frac{q^2}{m_{B_s}^2-m_\phi^2}\\ 
T_3(q^2)&=&\frac{i}{2}\left[A(q^2)-B(q^2)\right]+\frac{i}{2} H(q^2)  
\left( m_{B_s}^2-m_\phi^2 \right). 
\end{eqnarray} 
Again, the condition $T_1(0)=T_2(0)$ is manifestly satisfied. 
Now, the polar contributions from the $1^-$ and $1^+$ intermediate 
states have been computed in \cite{report} with the following 
results; if the $1^{-}$ state is taken into account: 
\begin{eqnarray} 
A(q^2)&=& \frac{i2\sqrt{2}\hat{F}\lambda g_V P}{(m_P^2-q^2)\sqrt{m_{B_s}}}\\ 
B(q^2)&=&\frac{-i2\sqrt{2}\hat{F}\lambda g_V 
m_{B_s}^{3/2}}{m_P^2-q^2}\\ 
H(q^2)&=&\frac{-i2\sqrt{2}\hat{F}\lambda g_V 
}{(m_P^2-q^2)\sqrt{m_{B_s}}}. 
\end{eqnarray} 
If instead we consider the $1^+$ contribution: 
\begin{eqnarray} 
A(q^2)&=& \frac{-i\sqrt{2m_{B_s}}\hat{F}^+g_V(\zeta-2\mu m_\phi 
)}{(m_P^2-q^2)}\\ B(q^2)&=&0\\ 
H(q^2)&=&\frac{-i2\sqrt{2m_{B_s}}\hat{F}^+\mu g_V 
}{(m_P^2-q^2)m_{B_s}}. 
\end{eqnarray} 
where $m_P$ is the mass of the intermediate polar state. Let us 
consider the contribution due to the $1^-$ state using the results 
for $A(q^2)$ and $B(q^2)$ obtained in \cite{report}. We find: 
\begin{equation} 
T_1^P(q^2=0)=-\frac{\hat{F}\lambda 
g_V}{\sqrt{2m_{B_s}}}=T_2^P(q^2=0), \label{strutt} 
\end{equation} 
neglecting a term $\frac{m_{B_s}^2 -m_\phi^2}{m_{B_s^{*2}}}$. The 
contribution to these form factors due to the $1^+$ state is 
sub-leading, being  $O(1/m_B)$. The form factor $T_3^P(0)$ has the 
same structure $(\ref{strutt})$ of $T_{1,2}^P(0)$ for the $1^-$ 
contribution. The sub-leading contribution from $1^+$ is instead: 
\begin{equation} 
T_3^P(0)=\left(\sqrt{\frac{m_{B_s}}{2}}(\hat{F}^+ g_V(\zeta-2\mu 
m_\phi))\frac{1}{m_P^2}+\frac{(m_{B_s}^2-m_\phi^2)}{m_P^2} 
\sqrt{\frac{2}{m_{B_s}}} (\hat{F}^+ g_V\mu) \right). 
\end{equation} 
We do not include the sub-leading contributions in the numerical 
analysis since they turn out to be very small corrections, 
certainly below the theoretical error induced by the model itself 
(varying e.g. the parameter $\Delta_H$ in the range $0.5,0.6,0.7$ 
GeV). 

\subsection{Numerical results for the form factors}
The form factors used in the branching ratio calculation are 
obtained adding up polar and direct contributions: 
\begin{equation} 
F(q^2)=F^D(q^2)+F^P(q^2)~.
\end{equation} 
The $q^2$ dependence of the form factors is given in Fig. 3 and
4. The dependence on the model parameter $\Delta_H$ is less than $10\%$.
Fig. 3 is similar to what obtained  for the decay $B \to \rho \ell \nu$
\cite{brholnu} (note however that  Fig.3 of \cite{brholnu} contains a misprint
and $A_1$ and $A_2$ are  interchanged). As in the $B \to \rho$
decay the $V$ form factor results in the model from a large cancellation
between the direct and polar term. Therefore the result for the $V$ form 
factor has a large incertitude.

In order to compare with results from other
approaches we consider the following two parameterizations of the form
factors:  
\begin{equation} 
F(q^2)=\frac{F(0)}{1~-~a_F \left(\frac{q^2}{m_{B_s}^2}\right) 
+~b_F \left(\frac{q^2}{m_{B_s}^2}\right)^2}~, 
\label{ffparam1} 
\end{equation}
and 
\begin{equation} 
F(q^2)=F(0){\mathrm exp} \left[ c_1 \frac{q^2}{m_{B_s}^2} 
+c_2\left( \frac{q^2}{m_{B_s}^2}\right)^2+c_3
\left( \frac{q^2}{m_{B_s}^2}\right)^3 \right] ~.  
\label{ffparam2} 
\end{equation} 
The coefficients $a_F$, $b_F$, $c_1$, $c_2$ and $c_3$ are given in 
Table~III. Results can be compared with those obtained from QCD
sum--rules \cite{ball} (see Table~III and Fig.~4 of that paper). In order to
allow an easier comparison we write their $F(0)_{\mathrm {SR}}$ in Table~III.
As already stated our $V$ form factor is affected by a large incertitude.
Concerning $A_0$, $A_1$, $A_2$, their value at $q^2=0$ is different in the two
models but their shape as a function of $q^2$ is quite similar. The tensor
form factors have a more pronounced pole-like behaviour in the QCD sum--rule
calculation than in the present one. 

\section{Relations between the form factors} 
Relations between the form factors can be established using the equations
of motion for the heavy quarks or taking limits of the general expressions
calculated in the preceding sections. They are useful to link different
decay processes and as a cross--check of the calculations. 
\subsection{Semileptonic and tensor currents}
The equations of motion of the heavy quark implies 
\be  
\frac{1+{\slash v}}{2}\; b=b \; , 
\ee 
and in the $b$ rest frame one has 
\be 
\gamma^0 b=b \; , 
\ee 
which can be used to relate vector and tensor currents \cite{isgurgamma} 
\be 
{\bar q}^a\sigma_{0i}(1+\gamma_5)Q=-i{\bar q}^a\gamma_i(1-\gamma_5)Q \; . 
\label{relat} 
\ee 
Therefore the form factors $T_1$, $T_2$, $T_3$ of eq. (\ref{param})  
can be related to those  describing the weak semileptonic transition  
$B \to \phi$ of eq. (\ref{ffsemilep}) or $B \to \rho$, using $SU(3)$  
symmetry. 
 
Using (\ref{relat}), the form factors $T_1$, $T_2$, $T_3$ 
are related to the form factors $V$, $A_1$ 
and $A_2$  as follows: 
\begin{eqnarray} 
T_1(q^2)&=& \frac{1}{2 m_{B_s}} \left[ \frac{m_{B_s}^2-m_\phi^2+q^2} 
{m_{B_s}+m_\phi} V(q^2) - (m_{B_s}+m_\phi) A_1(q^2) \right]\\ 
T_2(q^2)&=& \frac{1}{2 m_{B_s} (m_{B_s}^2-m_\phi^2)} \Big\{ \frac{V(q^2)} 
{m_{B_s}+m_\phi} \left[m_{B_s}^4-2m_{B_s}^2 (m_\phi^2+q^2) + 
(m_\phi^2-q^2)^2 \right] \nonumber \\ 
&-& A_1(q^2) (m_{B_s}+m_\phi) (m_{B_s}^2-m_\phi^2+q^2) 
\Big\}\\ 
T_3(q^2)&=& \frac{1}{2 m_{B_s} (m_{B_s}+m_\phi)} \Big\{ V(q^2)  
(m_{B_s}^2+3 m_\phi^2-q^2) + A_1(q^2) (m_{B_s}+m_\phi)^2  
\nonumber \\ 
&-& \frac{A_2(q^2)} 
{q^2} (m_{B_s}^2-m_\phi^2) (m_{B_s}^2-m_\phi^2+q^2)\Big\}. 
\end{eqnarray} 
In a similar way the other equivalent parameterization in terms of $A$,  
$B$ and $H$ is related to the form factors $V$, $A_1$ and $A_2$: 
\begin{eqnarray} 
A(q^2) &=& i\left\{\frac{q^2-M_B^2-m_{K^*}^2}{M_B} 
\frac{V(q^2)}{M_B+m_{K^*}}-\frac{M_B+m_{K^*}}{M_B}A_1(q^2)\right\} 
\label{aq2} \\ 
B(q^2) &=& i\frac{2M_B}{M_B+m_{K^*}} V(q^2)  \label{bq2} \\ 
H(q^2) &=& \frac{2 i}{M_B}\left\{ 
\frac{V(q^2)}{M_B+m_{K^*}}+\frac{1}{2q^2}\frac{q^2+M_B^2-m_{K^*}^2} 
{M_B+m_{K^*}}  A_2(q^2)\right\}~~. \label{hq2}  
\end{eqnarray} 
These relations are strictly valid only for $q^2\approx q^2_{max}$.  

\subsection{Final Hadron Large Energy Limit}
We examine a particular limit for the $B \to \phi$ semileptonic form factors,
the one of heavy mass for the initial meson and of large energy 
for the final one (LEET). The expressions of the form factors simplify in the
limit and for $B \to V l \nu$, they reduce only to two
independent functions \cite{leet}. The
four-momentum of  the heavy meson is written as $p= M_H v$ in terms of the
mass and the  velocity of the heavy meson. The four-momentum of the light
vector meson is written as $p'=E n$ where $E=v \cdot p'$ is the energy of the
light meson and $n$ is a four-vector defined by $v \cdot n=1, n^2=0$. 
The relation between $q^2$ and $E$ is:
\begin{equation}
q^2=M_H^2- 2 M_H E +m_V^2
\end{equation}
The large energy limit is defined as : 
\be
\Lambda_{QCD}, m_V << M_H, E
\ee
keeping $v$ and $n$ fixed and $m_V$ is, in our case, the mass of the 
$\phi$. The relations between the form factors appearing in the LEET limit 
constitute a powerful theoretical cross--check of the formulas derived
in the previous sections. Note that in our model the polar diagram of Fig. 2
is sub-leading with respect to the direct diagram of Fig. 1 in the LEET limit;
therefore only the direct part of the form factors contribute to the expression
of the LEET form factors. In agreement with the results obtained in
\cite{moriond} we find the following result: 
\ba
A_0(q^2)&=&\left(1-\frac{m_V^2}{M_H E}\right)\zeta_{||}(M_H,E)
+\frac{m_V}{M_H}\,\zeta_{\perp}(M_H,E) \label{fatta0}\\
A_1(q^2)&=&\frac{2E}{M_H + m_V}\,\zeta_{\perp}(M_H,E) \label{fatta}\\
A_2(q^2)&=&\left(1+\frac{m_V}{M_H}\right)\left[\zeta_{\perp}(M_H,E)-
\frac{m_V}{E}\zeta_{||}(M_H,E)\right]\\
V(q^2)&=&\left(1+\frac{m_V}{M_H}\right)\zeta_{\perp}(M_H,E).
\label{fattv}
\ea
The explicit expressions for $\zeta_{||}$ and $\zeta_{\perp}$
are as follows \cite{moriond}:
\ba
\zeta_{||}(M_H,E)&=&\frac{\sqrt{M_H Z_H}\; m_V^2}{2 E f_V} \Big[
I_3\left(\frac{m_V}{2}\right)-I_3\left(-\frac{m_V}{2}\right)
\nonumber \\
&+&4 \Delta_H m_V\; Z
\Big] \sim \frac{\sqrt{M_H}}{E} 
\label{zetapar}
\ea
\be
\zeta_{\perp}(M_H,E)=\frac{\sqrt{M_H Z_H}\; m_V^2} {2 E f_V}
\left[I_3(\Delta_H) + m_V^2 \; Z \right]
\sim \frac{\sqrt{M_H}}{E}\; ,
\label{zetaperp}
\ee
where terms proportional to the constituent light quark mass $m$ have been
neglected. Note that to obtain the correct results for the $\phi$
meson one has to replace:
\begin{equation}
\frac{m_V^2}{f_V} \to g_{\phi ss} \equiv \frac{m_\phi^2}{f_\phi} 
\frac{\sqrt{6}}{3 \cos (39.4^o)}
\label{gphiss}
\end{equation}
in order to take into account the constituent quark structure of the $\phi$
meson \cite{escribano}. The angle $39.4^o$ in (\ref{gphiss}) is the
$\omega$--$\phi$ mixing. It is interesting to note that in LEET one can also
relate the form factor $T_1$, $T_2$ and $T_3$ to the semileptonic ones and to
the $\zeta_{\perp}$ and $\zeta_{||}$ form factors of the LEET limit
\cite{leet}: 
\begin{eqnarray} 
T_1(q^2)&=&\zeta_{\perp}(M_H,E)\,,\\
T_2(q^2)&=&\left(1-\frac{q^2}{M_H^2-m_V^2}\right)\zeta_{\perp}(M_H,E)\,,\\
T_3(q^2)&=&\zeta_{\perp}(M_H,E)-\frac{m_V}{E}\left(1-\frac{m_V^2}{M_H^2}\right)
\zeta_{||}(M_H,E)\,.
\label{fattt3}
\end{eqnarray}
$\zeta_{\perp}$ and $\zeta_{||}$ obtained in this way agree with those of
(\ref{zetapar},\ref{zetaperp}).

\section{Decay distribution and asymmetry}
The dilepton invariant mass spectrum for the decay $B_s \to \phi \ell^+
\ell^-$ can be written in terms of the adimensional masses $\sh \equiv
q^2/m_{B_s}^2$, $\mlh \equiv m_\ell/m_{B_s}$ and $\mvh \equiv m_\phi/m_{B_s}$
\cite{ali}:
\begin{eqnarray}
\frac{\d \Gamma}{\d \sh} &=& \frac{G_F^2 \, \alpha^2 \, m_{B_s}^5}
{2^{10} \pi^5} \left| V_{ts}^\ast  V_{tb} \right|^2 \, \uh \times   
\Bigg\{\frac{|a|^2}{3} \sh \eta (1+2 \frac{\mlh^2}{\sh})
+|e|^2 \sh \frac{\uh^2}{3} \Bigg. \nonumber \\
&+& \Bigg. \frac{1}{4 \mvh^2} \left[ 
|b|^2 (\eta-\frac{\uh^2}{3} + 8 \mvh^2 (\sh+ 2 \mlh^2) ) 
+ |f|^2 (\eta -\frac{ \uh^2}{3} + 8 \mvh^2 (\sh- 4 \mlh^2)) 
\right] \Bigg. \nonumber \\
&+&\Bigg. \frac{\eta}{4 \mvh^2} \left[ |c|^2 (\eta - \frac{\uh^2}{3}) 
+ |g|^2 \left(\eta -\frac{\uh^2}{3}+4 \mlh^2(2+2 \mvh^2-\sh) \right) 
\right] \Bigg. \nonumber \\
&-& \Bigg. \frac{1}{2 \mvh^2}
\left[ {\rm Re}(bc^\ast) (\eta -\frac{ \uh^2}{3})(1 - \mvh^2 - \sh) 
+{\rm Re}(fg^\ast) ((\eta -\frac{ \uh^2}{3})\
(1-\mvh^2 -\sh) +4\mlh^2 \eta) \right] \Bigg.\nonumber \\
&-& \Bigg. 2 \frac{\mlh^2}{\mvh^2} \eta  \left[ {\rm Re}(fh^\ast)-
{\rm Re}(gh^\ast) (1-\mvh^2) \right] +\frac{\mlh^2}{\mvh^2} \sh \eta |h|^2
\Bigg\}
\label{dgammall}
\end{eqnarray}
where 
\begin{equation}
\eta = 1+\frac{m_\phi^4}{m_{B_s}^4}+\frac{q^4}{m_{B_s}^4}
-2 \frac{m_\phi^2}{m_{B_s}^2}(1+\frac{q^2}{m_{B_s}^2}) 
-2 \frac{q^2}{m_{B_s}^2}
\end{equation}
and 
\begin{equation}
\uh = \sqrt{\eta\; \left(1-4 \frac{m_{\ell}^2}{q^2}\right)}.
\end{equation}
The functions $a$ to $h$ contain the form factors dependence and the Wilson
coefficients (see the Appendix). The invariant muon mass distribution for the
decay $B_s \to \phi \mu^+ \mu^-$ is given in Fig. 5.
Integrating the differential decay distribution (\ref{dgammall}) allows to
compute the branching fraction for the $B_s \to \phi \mu^+ \mu^-$ decay:
\begin{equation}
{\cal B} (B_s \to \phi \mu^+ \mu^-) = 8.8 \times 10^{-5}\; .
\label{branching}
\end{equation}
Note however that this number is model dependent not only due to the form
factors but also to the way $c\bar{c}$ resonances are taken into account
(see formula (\ref{nonpertcc}) in the Appendix).
For a comparison we calculated the same branching fraction excluding the effect
of the $c\bar{c}$ resonances:
\begin{equation}
{\cal B} (B_s \to \phi \mu^+ \mu^-)_{\mathrm non-resonant} = 2.5 \times
10^{-6}\; . 
\label{branchingnp}
\end{equation}
This amounts to use Eq. (\ref{ypert}) instead of Eq. (\ref{nonpertcc}) for 
the calculation. Finally in order to have a more realistic estimate of the
branching ratio we use the complete expression (\ref{nonpertcc}) but exclude
the $c\bar{c}$ resonance regions 2.9--3.3 GeV and 3.6--3.8 GeV from the
integration as in \cite{cdf}:
\begin{equation}
{\cal B} (B_s \to \phi \mu^+ \mu^-)_{\mathrm exp-like} = 1.9 \times
10^{-6}\; . 
\label{branchingexplike}
\end{equation}

The differential forward--backward asymmetry is given by \cite{morozumi}
\begin{equation}
\frac{d {\cal A}_{\rm FB}}{d \sh} = 
-\int_0^{\uh(\sh)} d\uh \frac{d^2\Gamma}{d\uh d\sh}
+ \int_{-\uh(\sh)}^0 d\uh \frac{d^2\Gamma}{d\uh d\sh} \; .
\label{fba}
\end{equation}
For $B_s\to \phi \ell^+\ell^-$ decays we obtain:
\begin{equation}
\frac{d {\cal A}_{\rm FB}}{d \sh} =
\frac{G_F^2 \, \alpha^2 \, m_{B_s}^5}{2^{10} \pi^5} 
\left| V_{ts}^\ast V_{tb} \right|^2 \, \sh \uh(\sh)^2
\left[ {\rm Re}(be^\ast) + {\rm Re}(af^\ast) \right]\; .
\end{equation}
In Fig. 6 we plot the differential forward--backward asymmetry
normalized to the differential decay rate:
\begin{equation}
\frac{d \bar{\cal A}_{\rm FB}}{d \sh} = 
\frac{d {\cal A}_{\rm FB}}{d \sh}/\frac{d \Gamma}{d\sh}~.
\end{equation}
The position of the zero $\sh_0$ is given by
\begin{equation}
{\rm Re}(\cne(\sh_0)) =- \frac{\mbh}{\sh_0} \cse 
\left\{\frac{T_2(\sh_0)}{A_1(\sh_0)} (1-\mvh)+
\frac{T_1(\sh_0)}{V(\sh_0)} (1+\mvh)\right\} \; ,
\label{zero}
\end{equation}
Note that in LEET the ratios $T_1/V$ and $T_2/A_1$ are
simple functions of $\mvh$ and $\sh$ with no hadronic uncertainties,
as can be seen from formulas (\ref{fatta0}-\ref{fattt3}). For a detailed study
including radiative corrections see \cite{thorsten}. Decays such as
$B_s\to\phi\ell^+\ell^-$ involve the Wilson  coefficients $\cse$, $\cne$ and
$C_{10}$. In extensions of the SM they can assume rather different values from
those expected in SM. In particular the position of the zero in the
forward-backward asymmetry is a measure of $C_7/{\mathrm Re} (C_9)$ which, as
shown above, depends on form factor ratios. This decreases the model
dependence of this number. Moreover the sign of $C_7$ can be opposite to the
SM one in beyond-SM scenarios. All these elements explain the relevance of the
experimental study of the forward-backward asymmetry and the need of form
factors computations, despite of their model-dependent nature. 

\section{Conclusions}
The exclusive process $B_s\to\phi\mu^+\mu^-$ belongs to a set of
processes, like $B\to K^*\mu^+\mu^-$, $B\to K\mu^+\mu^-$, $B_s\to
\eta\mu^+\mu^-$, which will be accurately studied at
$B$-factories. In this paper we have examined
$B_s\to\phi\mu^+\mu^-$ in the framework of a Constituent Quark
Meson Model. The $\phi$ meson is coupled using the VMD hypothesis. The model
has extensively been tested in a number of exclusive processes \cite{nc}. It
provides results in good agreement with experimental data and with those
obtained using QCD Sum Rules. In order to have a better understanding and a
check of the form factors, we have considered the LEET limit of the
$B_s\to\phi\mu^+\mu^-$ decay obtaining consistency among $V$,$A$ and $T$
direct form factors. We have studied the decay distribution and the
forward-backward asymmetry. As shown in \cite{jhep}, CQM offers a versatile 
calculation framework also to study more exotic processes involving higher
spin meson states. This aspect of the model will be very useful for the study
of higher spin $B$ and $B_s$ rare meson decays as soon as data on these states
will become available.

\section*{Appendix} 
\subsection{Integrals}
We list the explicit expressions for the integrals used in the 
text: 
 
\begin{eqnarray} 
 \Omega_1&=&\frac{ I_3(-x/2)-I_3(x/2)+\omega[I_3(\Delta_1)- 
I_3(\Delta_2)]}{2 x (1-\omega^2)} - \frac{[\Delta_1-\omega x/2]Z} 
{1-\omega^2} 
 \\ 
\Omega_2&=&\frac{ 
-I_3(\Delta_1)+I_3(\Delta_2)-\omega[I_3(-x/2)-I_3(x/2)]} {2 x 
(1-\omega^2)} - \frac{[x/2- \Delta_1\omega ]Z}{1-\omega^2} 
 \\ 
\Omega_3&=&\frac{K_1}{2}+\frac{2 \omega 
K_4-K_2-K_3}{2(1-\omega^2)} 
 \\ 
\Omega_4&=&\frac{-K_1}{2(1-\omega^2) }+\frac{3 K_2-6 \omega 
K_4+K_3 (2\omega^2+1)}{2(1-\omega^2)^2} 
 \\ 
\Omega_5&=&\frac{-K_1}{2(1-\omega^2) }+\frac{3 K_3-6 \omega 
K_4+K_2 (2\omega^2+1)}{2(1-\omega^2)^2} 
 \\ 
\Omega_6&=&\frac{K_1\omega}{2(1-\omega^2) }+\frac{2 
K_4(2\omega^2+1)- 3\omega( K_2+K_3) }{2(1-\omega^2)^2}\\ Z &=& 
\frac{iN_c}{16\pi^4} \int^{\mathrm {reg}} 
\frac{d^4k}{(k^2-m^2)[(k+q)^2-m^2](v\cdot k + \Delta_1 + 
i\epsilon)}\nonumber \\ &=&\frac{I_5(\Delta_1, 
x/2,\omega)-I_5(\Delta_2,- x/2,\omega)}{2 x}, 
\end{eqnarray} 
where $x=m_\phi$, $\omega$, $\Delta_1$, $\Delta_2$, are
specified in the text and $K_i$'s are given by: 
\begin{eqnarray} 
K_1&=&m^2 Z -I_3(\Delta_2) \\ K_2&=&\Delta_1^2 Z -\frac{I_3(x/2)- 
I_3(-x/2)}{4 x}[\omega ~ x + 2 \Delta_1] 
 \\ 
K_3&=&\frac{x^2}{4} Z +\frac{I_3(\Delta_1)-3 
I_3(\Delta_2)}{4}+\frac{\omega}{4}[\Delta_1 I_3(\Delta_1)- 
\Delta_2 I_3(\Delta_2)] 
 \\ 
K_4&=&\frac{x \Delta_1}{2} Z +\frac{\Delta_1[I_3(\Delta_1)- 
I_3(\Delta_2)]}{2 x}+\frac{I_3(x/2)-I_3(-x/2)}{4}, 
\end{eqnarray} 
expressed in terms of the $I_i$ integrals, regularized using the 
Schwinger's proper time regularization method: 
\begin{eqnarray} 
I_1&=&\frac{iN_c}{16\pi^4} \int^{reg} \frac{d^4k}{(k^2 - m^2)} 
={{N_c m^2}\over {16 \pi^2}} \Gamma\left(-1,{{{m^2}} \over 
{{{\Lambda}^2}}},{{{m^2}}\over {{{\mu }^2}}}\right) 
\\ 
I_3(\Delta) &=& - \frac{iN_c}{16\pi^4} \int^{\mathrm {reg}} 
\frac{d^4k}{(k^2-m^2)(v\cdot k + \Delta + i\epsilon)}\nonumber \\ 
&=&{N_c \over {16\,{{\pi }^{{3/2}}}}} 
\int_{1/{{\Lambda}^2}}^{1/{{\mu }^2}} {ds \over {s^{3/2}}} \; e^{- 
s( {m^2} - {{\Delta }^2} ) }\; \left( 1 + {\mathrm {erf}} 
(\Delta\sqrt{s}) \right)\\ I_5(\Delta_1,\Delta_2,\omega) &= & 
\frac{iN_c}{16\pi^4} \int^{\mathrm {reg}} 
\frac{d^4k}{(k^2-m^2)(v\cdot k + \Delta_1 + i\epsilon ) (v'\cdot k 
+ \Delta_2 + i\epsilon )} \nonumber \\ 
 & = & \int_{0}^{1} dx \frac{1}{1+2x^2 (1-\omega)+2x 
(\omega-1)}\times\nonumber\\ &&\Big[ 
\frac{6}{16\pi^{3/2}}\int_{1/\Lambda^2}^{1/\mu^2} ds~\sigma \; 
e^{-s(m^2-\sigma^2)} \; s^{-1/2}\; (1+ {\mathrm {erf}} 
(\sigma\sqrt{s})) +\nonumber\\ 
&&\frac{6}{16\pi^2}\int_{1/\Lambda^2}^{1/\mu^2} ds \; 
e^{-s(m^2-2\sigma^2)}\; s^{-1}\Big]. 
\end{eqnarray} 
Here we have used the definitions: 
\begin{eqnarray} 
\Gamma(\alpha,x_0,x_1) &=& \int_{x_0}^{x_1} dt\;  e^{-t}\; 
t^{\alpha-1}\\ {\rm erf}(z) &=& \frac{2}{\sqrt{\pi}}\int_{0}^{z}dx 
e^{-x^2}\\ 
\sigma(x,\Delta_1,\Delta_2,\omega)&=&{{{\Delta_1}\,\left( 1 - x 
\right)  + {\Delta_2}\,x}\over {{\sqrt{1 + 2\,\left(\omega -1 
\right) \,x + 2\,\left(1-\omega\right) \,{x^2}}}}}.\\ 
\end{eqnarray} 

\subsection{Integrals in the LEET limit}
We list the expressions for the integrals used to compute the form
factors in the LEET limit: 
\begin{eqnarray} 
\Omega_1^{\mathrm LEET}&=&-\frac{1}{2 E} \left[I_3(\Delta_1) +x^2 Z \right]\\ 
\Omega_2^{\mathrm LEET}&=&-\frac{1}{2 E} \left[I_3(x/2) -I_3(-x/2)
+ 2x \Delta_1 Z \right]\\ 
\Omega_3^{\mathrm LEET}&=&-\frac{1}{8 E} \left[x (I_3(x/2) -I_3(-x/2)) 
+ 2 \Delta_1 (I_3(\Delta_1) +2 x^2 Z )
\right]\\ 
\Omega_4^{\mathrm LEET}&=& \frac{1}{2E} \Delta_1 I_3(\Delta_1)\\ 
\Omega_5^{\mathrm LEET}&=&\frac{x}{4 E} \left[I_3(x/2) -I_3(-x/2) \right]\\ 
\Omega_6^{\mathrm LEET}&=& \frac{x}{8 E^2} \left[x (I_3(x/2) -I_3(-x/2)) 
+ 2 \Delta_1 (I_3(\Delta_1) +4 x^2 Z )
\right]\; ,
\end{eqnarray}
where $\Delta_1=\Delta_H$, $x=m_\phi$.

\subsection{Auxiliary functions}
The auxiliary functions introduced in the formula for the invariant mass
spectrum for the $B_s \to \phi \ell^+ \ell^-$ decay are defined as 
\begin{eqnarray}
a(\sh) & = & \frac{2}{1 + \mvh} \cne(\sh) V(\sh) 
+ \frac{4 (\mbh + \msh)}{\sh} \cse T_1(\sh) \; , \\ 
b(\sh) & = & (1 + \mvh) \left[ \cne(\sh) A_1(\sh) 
+ \frac{2 (\mbh-\msh)}{\sh} (1 - \mvh) \cse T_2(\sh) \right] \; , \\
c(\sh) & = & \frac{1}{1 - \mvh^2} \left[ 
(1 - \mvh) \cne(\sh) A_2(\sh) 
+ 2 (\mbh-\msh) \cse \left( T_3(\sh) + \frac{1 - \mvh^2}{\sh} T_2(\sh) \right)
\right] \; , \\   
d(\sh) & = & \frac{1}{\sh} \left[ \cne(\sh) \left(
(1 + \mvh) A_1(\sh) - (1 - \mvh) A_2(\sh) - 2 \mvh A_0(\sh) \right) 
- 2 (\mbh-\msh) \cse T_3(\sh) \right] \; , \\
e(\sh) & = & \frac{2}{1 + \mvh} \ct V(\sh) \; , \\
f(\sh) & = & (1 + \mvh) \ct A_1(\sh) \; , \\
g(\sh) & = & \frac{1}{1 + \mvh} \ct A_2(\sh) \; , \\
h(\sh) & = & \frac{1}{\sh} \ct \left[
(1 + \mvh) A_1(\sh) - (1 - \mvh) A_2(\sh) - 2 \mvh A_0(\sh) \right] .
\end{eqnarray}
The Wilson-coefficients $C_i$ are those of \cite{buras} (see Table
I). The effective coefficients $C^{\rm eff}_i$ are defined as
follows  \begin{equation}
C_7^{\rm eff} \equiv C_7 -C_5/3 -C_6
\end{equation}
\begin{equation}
\cne (\sh) = C_9 +  \Upsilon(\sh ) \; ,
\label{c9eff}
\end{equation}
where $\Upsilon(\sh)$ is matrix elements of the four-quark operators
(for a detailed discussion on the perturbative and non-perturbative
contributions see \cite{ali}). A perturbative calculation yields \cite{munz}:
\begin{eqnarray} 
{\Upsilon}_{\rm pert} (\sh) & = & \aleph(z,\sh)
\left(3 C_1 + C_2 + 3 C_3 + C_4 + 3 C_5 + C_6 \right)
\nonumber \\
&-& \frac{1}{2} \aleph(1,\sh) \left( 4 C_3 + 4 C_4 + 3 C_5 + C_6 \right) 
-\frac{1}{2} \aleph(0,\sh) \left( C_3+ 3 C_4 \right) \nonumber \\
&+& \frac{2}{9} \left( 3 C_3 + C_4 +3 C_5 + C_6 \right) \; , 
\label{ypert}
\end{eqnarray}
where $z \equiv m_c/m_b$ and 
\begin{eqnarray}
\aleph(z,\sh) &=& -\frac{8}{9}\ln\frac{m_b}{\mu} - \frac{8}{9}\ln z +
\frac{8}{27} + \frac{4}{9} x -\frac{2}{9} (2+x) |1-x|^{1/2} \left\{
\matrix{
\left( \ln\left| \frac{\sqrt{1-x} + 1}{\sqrt{1-x} - 1}\right| - i\pi
\right), & x < 1 \cr
2 \arctan \frac{1}{\sqrt{x-1}}, &  x > 1\cr}
\right. \\
\aleph(0, \sh) & = & -\frac{8}{9} \ln\frac{m_b}{\mu} +\frac{8}{27}  -
\frac{4}{9} \ln \sh + \frac{4}{9} i\pi 
\end{eqnarray}
and $x \equiv 4z^2/\sh$.
In order to incorporate the non-perturbative contributions to $\Upsilon$ we
follow the phenomenological prescription of \cite{morozumi} where 
$c\bar c$ resonance contributions from $J/\psi, \psi^\prime,\dots$
parametrized in the form of a Breit-Wigner, are added to the perturbative 
result:
\begin{equation}
\Upsilon(\sh) = \Upsilon_{\rm pert}(\sh) + \frac{3 \pi}{\alpha^2} \left(
3 C_1 + C_2 + 3 C_3 + C_4 + 3 C_5 + C_6 \right)
\sum_{V_i = \psi(1s),..., \psi(6s)} \kappa_i
\frac{\Gamma(V_i \rightarrow \ell^+ \ell^-)\, m_{V_i}}{
{m_{V_i}}^2 - q^2 - i m_{V_i} \Gamma_{V_i}}~.
\label{nonpertcc}
\end{equation}
The numerical values used for the masses, widths and branching fractions
of the $1^{--}$ charmonium resonances are given in Table V
(data from\cite{pdg}).
The factors $\kappa_i$ correct for the naive factorization 
approximation. $\kappa_1$ is calculated comparing the experimental 
$B_s \to J/\psi(1S) \phi$ branching fraction to the one predicted by
our calculation using the $J/\psi \to  \mu^+ \mu^-$ experimental branching
fraction:
\begin{equation}
{\cal B} (B_s \to \phi J/\psi \to \phi \mu^+ \mu^-)
={\cal B} (B_s \to \phi J/\psi)_{\mathrm exp} 
{\cal B}(J/\psi\to \mu^+ \mu^-)_{\mathrm exp}~. 
\end{equation}
The $q^2$ integration range for the calculated branching is
$(m_{J/\psi}-\Gamma_{J/\psi})^2 <q^2< (m_{J/\psi}+\Gamma_{J/\psi})^2$,
which is a $2 \Gamma_{J/\psi}$ interval around the $J/\psi$ resonance. The
result is that a $\kappa_1=1.36$ is needed to correct for the 
factorization result. By taking $4 \Gamma_{J/\psi}$ for the integration
interval around the $J/\psi$ resonance only changes the branching ratio ${\cal
B}(B_s \to J/\psi \phi)$ from $9$ to $11 \times 10^{-4}$ which is within the
experimental error. For the other $\kappa_i$ values we take again $1.36$ as no
experimental values are known for the higher $1^{--}$ charmonium resonances.
Note that for the $SU(3)$ related decay $B \to J/\psi K^*$ the $\kappa$ factor
is estimated to be $\simeq 2.3$ \cite{ligeti} using inclusive $B \to X_s
\ell^+ \ell^-$ data. However using exclusive data a smaller $\kappa \simeq
1.6$ is obtained \cite{ali}.

\acknowledgements ADP acknowledges support  from EU-TMR programme, 
contract CT98-0169. He is also grateful to J.O. Eeg and A. Hiorth 
for their hospitality at the University of Oslo.

\section*{Tables}
\begin{table}
\begin{center}
\vbox{\offinterlineskip
\halign{&#& \strut\quad#\hfil\quad\cr
\tableline
\tableline
& && && && && && && && && &\cr
&$C_1$ && $C_2$ && $C_3$ && $C_4$ && $C_5$ && $C_6$ && 
$C_7^{\rm eff}$ && $C_9$ && $C_{10}$ &\cr
& && && && && && && && && &\cr
\tableline
& && && && && && && && && &\cr
&$-0.248$ && $+1.107$ && $+0.011$ && $-0.026$ && $+0.007$ && $-0.031$ &&
$-0.313$ && $+4.344$ && $-4.669$&\cr 
& && && && && && && && && &\cr
\tableline
\tableline
& && && && && && && && && &\cr}}
\caption{Wilson coefficients used in the numerical
calculations. $C_7^{\rm eff} \equiv C_7 -C_5/3 -C_6$.} 
\end{center} 
\label{tab:wc}
\end{table}

\begin{table}
\vbox{\offinterlineskip
\halign{&#& \strut\quad#\hfil\quad\cr
\tableline
\tableline
& && &\cr
& $m_W$ && 80.41 GeV &\cr   
& $m_Z$ && 91.1867 GeV &\cr
& $\sin^2 \theta_W $ && 0.2233 &\cr
& $\mu$ (IR cutoff) && 0.51 GeV &\cr
& $\Lambda$ (UV cutoff) && 1.25 GeV &\cr
& $\Delta_H$ && 0.6 GeV &\cr
& $m_s$ (constituent) && 0.51 GeV &\cr
& $m_\phi$ && 1.02 GeV &\cr
& $f_\phi$ && 0.2491 GeV$^2$ &\cr
& $m_c$ && 1.25 GeV &\cr
& $m_b$ && 4.8 GeV &\cr
& $m_t$ && 173.8 GeV &\cr
& $\mu$ (scale) && $m_b$ &\cr
& $1/\alpha_{\mathrm em}$ && 129 &\cr
& $\alpha_s (m_Z) $ && 0.119 &\cr
& $|V^*_{ts} V_{tb}|$ && 0.04022 &\cr
& $|V^*_{ts} V_{tb}|/|V_{cb}|$ && 1 &\cr
& && &\cr
\tableline
\tableline
& && &\cr}}
\begin{minipage}{.35\linewidth}
\caption{Values of the parameters used in the numerical
calculations.} 
\end{minipage}
\label{tab:values}
\end{table}

\begin{table}
\vbox{\offinterlineskip
\halign{&#& \strut\quad#\hfil\quad\cr
\tableline
\tableline
& && && && && && && && &\cr
& && $V$ && $A_1$ && $A_2$ && $A_0$ && $T_1$ && $T_2$ && $T_3$ &\cr
& && && && && && && && &\cr
\tableline
& && && && && && && && &\cr
&$F(0)$ && $0.20$ && $0.59$ && $0.73$ && $0.29$ && $0.42$ &&
$0.42$ && $0.36$ &\cr 
& && && && && && && && &\cr
\tableline
& && && && && && && && &\cr
&$F(0)_{\mathrm {SR}}$ && $0.43$ && $0.30$ && $0.26$ && $0.38$ && $0.35$ &&
$0.35$ && $0.25$ &\cr 
& && && && && && && && &\cr
\tableline
& && && && && && && && &\cr
&$a_F$ && $+0.65$ && $-0.11$ && $+0.78$ && $+2.70$ && $+0.78$ &&
$+0.85$ && $+0.62$ &\cr 
&$b_F$ && $+0.96$ && $+0.49$ && $-0.52$ && $+2.30$ && $+0.07$ &&
$+12.9$ && $-0.88$ &\cr
& && && && && && && && &\cr
\tableline 
& && && && && && && && &\cr
&$c_1$ && $+0.41$ && $-0.19$ && $+1.03$ && $+3.65$ && $+0.75$ &&
$-1.89$ && $+0.48$ &\cr 
&$c_2$ && $+0.73$ && $+0.01$ && $-0.85$ && $-2.15$ && $+0.34$ &&
$+8.41$ && $+1.43$ &\cr 
&$c_3$ && $-2.47$ && $-0.61$ && $+3.47$ && $+1.60$ && $~0$ &&
$-25.3$ && $+1.22$ &\cr 
& && && && && && && && &\cr
\tableline
\tableline
& && && && && && && && &\cr}}
\begin{minipage}{.75\linewidth}
\caption{Values of the parameters of Eq. (\protect{\ref{ffparam1}})
and Eq. (\protect{\ref{ffparam2}}) for the $B\to \phi$ form factors in the
constituent quark model using the central value $\Delta_H=0.6$ GeV. Varying
$\Delta_H$ by $\pm 100$ MeV gives a variation of $10\%$ or less in the values
of the form factors. $F(0)_{\mathrm {SR}}$ is the form factor value in zero
from the QCD sum--rules.} 
\end{minipage} \label{tab:ffc}
\end{table}

\begin{table}
\vbox{\offinterlineskip
\halign{&#& \strut\quad#\hfil\quad\cr
\tableline
\tableline
& && &\cr
& Form Factor && Pole mass [GeV]&\cr
& && &\cr
\tableline
& && &\cr  
& $V$ && 5.416 &\cr
& $A_1$ && 5.75~ &\cr
& $A_2$ && 5.75~ &\cr
& $A_0$ && 5.75~ &\cr
& $T_1$ && 5.416 &\cr
& $T_2$ && 5.416 &\cr
& $T_3$ && 5.416 &\cr
& && &\cr
\tableline
\tableline
& && &\cr}}
\begin{minipage}{.3\linewidth}
\caption{Values of the mass poles for the polar form factors.}
\end{minipage}
\label{tab:poles}
\end{table}

\begin{table}
\vbox{\offinterlineskip
\halign{&#& \strut\quad#\hfil\quad\cr
\tableline
\tableline
& && && && &\cr
&  &&$M_{\psi{(nS)}}\,$[GeV]  &&  $\Gamma_{\psi{(nS)}}\,$[GeV]  &&
${\rm Br}(\psi{(nS)}\to\ell\,\bar\ell)$  &\cr 
& && && && &\cr
\tableline
& && && && &\cr
&$J/\psi$ && 3.097 && $8.7\times 10^{-5}$ && $5.88\times 10^{-2}$&\cr  
&$\psi{(2S)}$ && 3.686 && $2.77\times10^{-4}$ && $1.03\times 10^{-2}$&\cr  
&$\psi{(3S)}$ && 3.77 && $2.36\times 10^{-2}$ && $1.1\times10^{-5}$ &\cr 
&$\psi{(4S)}$ && 4.04 && $5.2\times 10^{-2}$ && $1.4\times 10^{-5}$ &\cr
&$\psi{(5S)}$ && 4.16 && $7.8\times 10^{-2}$ && $1.0\times 10^{-5}$ &\cr
&$\psi{(6S)}$ && 4.42 && $4.3\times 10^{-2}$ && $1.1\times 10^{-5}$ &\cr
& && && && &\cr
\tableline
\tableline
& && && && &\cr}} 
\begin{minipage}{.6\linewidth}
\caption{Masses, widths and leptonic branching ratios of the $1^{--}$ $c\bar c$
resonances. For $J/\psi$ and $\psi{(2S)}$ the branching ratio is the one to
$\mu^+ \mu^-$. For the other resonances is $\psi (nS) \to e^+ e^-$.} 
\end{minipage}
\label{tab:charm}
\end{table}

\newpage
\twocolumn
\section*{Figures}
\begin{figure}
\epsfysize=3truecm
\begin{center}
\epsffile{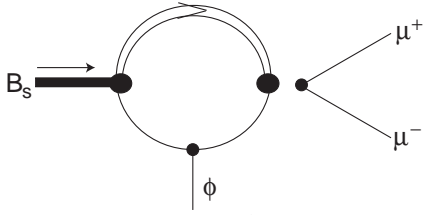}
\noindent
\caption{\footnotesize {\it Direct} diagram 
with $V,A,T$ currents changing $b\to s$. The momentum carried by 
the heavy quark is $m_Q v +\ell+k$ where $k$, the residual 
momentum $k\simeq \Lambda_{QCD}$, is due to the interaction of the 
heavy quark with the light degrees of freedom.} 
\end{center}
\label{fig:1} 
\end{figure}

\begin{figure}
\epsfysize=3truecm
\begin{center} 
\epsffile{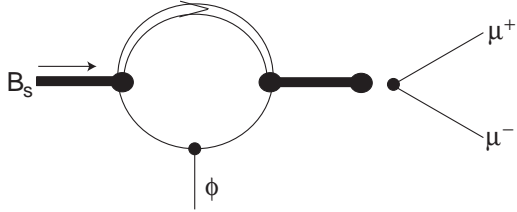} 
\caption{\footnotesize {\it Polar} diagram 
from $1^-$ and $1^+$ intermediate states.} 
\end{center} 
\label{fig:2} 
\end{figure} 

\begin{figure}
\epsfysize=4truecm
\begin{center} 
\epsffile{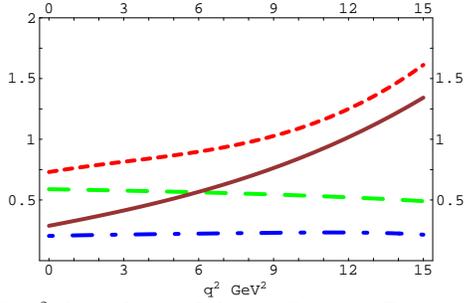} 
\caption{\footnotesize $q^2$ dependence of the semileptonic $B_s \to \phi$
form factors $V$ (dash-dotted), $A_1$ (large dashes), $A_2$ (small dashes)
and $A_0$ (continuous line).}  \end{center} 
\label{fig:3} 
\end{figure} 
 
\begin{figure}
\epsfysize=4truecm
\begin{center} 
\epsffile{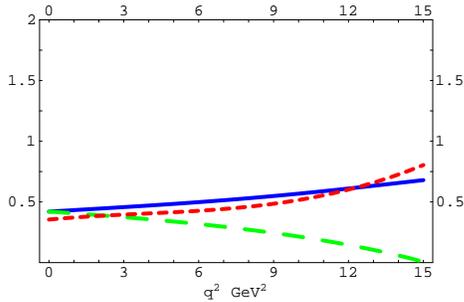} 
\caption{\footnotesize $q^2$ dependence of the tensor $B_s \to \phi$
form factors $T_1$ (continuous line), $T_2$ (large dashes)
and $T_3$ (small dashes).}  
\end{center} 
\label{fig:4} 
\end{figure}  

\begin{figure}
\epsfysize=4truecm
\begin{center} 
\epsffile{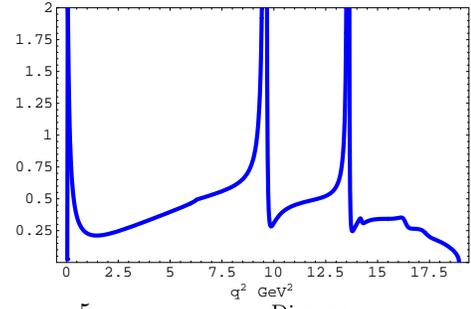}
\caption{\footnotesize Dimuon invariant mass distribution of the $B_s \to
\phi \mu^+ \mu^-$ decay. The vertical axis is in units of $10^{-5}$
GeV$^{-2}$.}
\end{center} 
\label{fig:5} 
\end{figure}  

\begin{figure}
\epsfysize=4truecm
\begin{center} 
\epsffile{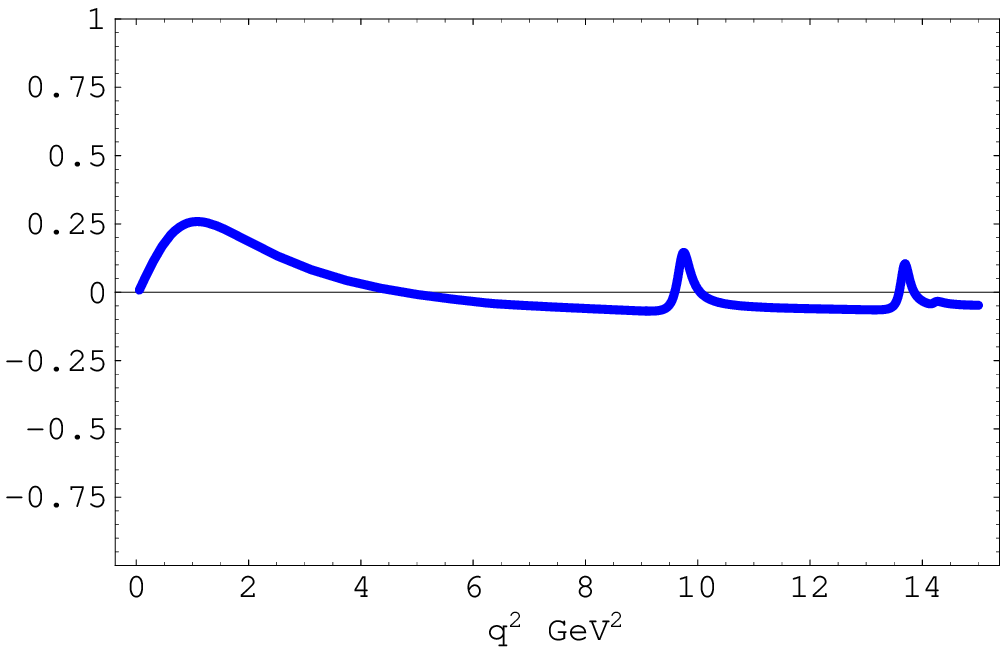} 
\caption{\footnotesize Normalized forward--backward asymmetry of the $B_s \to
\phi \mu^+ \mu^-$ decay. The two peaks at high $q^2$ values are due to the
$J/\psi$ and $\psi(2S)$ resonances.}  
\end{center} 
\label{fig:6} 
\end{figure} 

\begin{thebibliography}{99} 

\bibitem{nc}
A.~Deandrea, N.~Di Bartolomeo, R.~Gatto, G.~Nardulli and A.~D.~Polosa,
Phys.\ Rev.\ D {\bf 58}, 034004 (1998) [hep-ph/9802308];
for a review of the model see: A.~D.~Polosa, Riv.\ Nuovo\ Cim.{\bf23} N11,1
(2000) [hep-ph/0004183].

\bibitem{qmi}
D.~Ebert, T.~Feldmann, R.~Friedrich and H.~Reinhardt,
Nucl.\ Phys.\ B {\bf 434}, 619 (1995)
[hep-ph/9406220];
D.~Ebert, T.~Feldmann and H.~Reinhardt, Phys.\ Lett.\ B {\bf 388}, 154 (1996)
[hep-ph/9608223].

\bibitem{sanda}  
C.~S.~Kim, T.~Morozumi and A.~I.~Sanda,
Phys.\ Rev.\ D {\bf 56}, 7240 (1997) [hep-ph/9708299];
A.~Ali and G.~Hiller,
Eur.\ Phys.\ J.\ C {\bf 8}, 619 (1999) [hep-ph/9812267].

\bibitem{ball} 
P.~Ball and V.~M.~Braun,
Phys.\ Rev.\ D {\bf 58}, 094016 (1998)
[hep-ph/9805422].

\bibitem{buras} 
A.~J.~Buras, M.~Misiak, M.~Munz and S.~Pokorski,
Nucl.\ Phys.\ B {\bf 424}, 374 (1994)
[hep-ph/9311345].

\bibitem{Ebert:1983pk}
D.~Ebert and M.~K.~Volkov, Z.\ Phys.\ C {\bf 16}, 205 (1983).

\bibitem{ultimo}  
A.~Deandrea, R.~Gatto, G.~Nardulli, A.~D.~Polosa and N.~A.~Tornqvist,
Phys.\ Lett.\ B {\bf 502}, 79 (2001)
[hep-ph/0012120].

\bibitem{report}  
R.~Casalbuoni, A.~Deandrea, N.~Di Bartolomeo, R.~Gatto, F.~Feruglio 
and G.~Nardulli, Phys.\ Rept.\  {\bf 281}, 145 (1997) [hep-ph/9605342].

\bibitem{matsuki}  
T.~Matsuki and T.~Morii,
Phys.\ Rev.\ D {\bf 56}, 5646 (1997) 
[hep-ph/9702366].

\bibitem{wise} 
A.~V.~Manohar and M.~B.~Wise, ``Heavy quark physics,''
{\it  Cambridge Monographs on Particle Physics, Nuclear Physics, 
and Cosmology, Vol. 10} (2000).

\bibitem{brholnu}
A.~Deandrea, R.~Gatto, G.~Nardulli and A.~D.~Polosa,
Phys.\ Rev.\ D {\bf 59}, 074012 (1999) [hep-ph/9811259].

\bibitem{isgurgamma} 
N.~Isgur and M.~B.~Wise, Phys.\ Rev.\ D {\bf 42}, 2388 (1990).

\bibitem{leet} 
J.~Charles, A.~Le Yaouanc, L.~Oliver, O.~Pene and J.~C.~Raynal,
Phys.\ Rev.\ D {\bf 60}, 014001 (1999) [hep-ph/9812358];
Phys.\ Lett.\ B {\bf 451}, 187 (1999) [hep-ph/9901378].

\bibitem{moriond}
A.~Deandrea, 34th Rencontres de Moriond, QCD and Hadronic interactions, Les
Arcs, France, 20-27 Mar 1999, hep-ph/9905355.

\bibitem{escribano}
R.~Escribano and J.~M.~Frere,
Phys.\ Lett.\ B {\bf 459}, 288 (1999) [hep-ph/9901405].

\bibitem{ali}  
A.~Ali, P.~Ball, L.~T.~Handoko and G.~Hiller,
Phys.\ Rev.\ D {\bf 61}, 074024 (2000)
[hep-ph/9910221].

\bibitem{cdf}
T.~Affolder {\it et al.}  [CDF Collaboration],
Phys.\ Rev.\ Lett.\  {\bf 83}, 3378 (1999) [hep-ex/9905004].

\bibitem{morozumi}
A.~Ali, T.~Mannel and T.~Morozumi, Phys.\ Lett.\ B {\bf 273}, 505 (1991).

\bibitem{jhep}
A.~Deandrea, R.~Gatto, G.~Nardulli and A.~D.~Polosa,
JHEP {\bf 9902}, 021 (1999)
[hep-ph/9901266].

\bibitem{munz} 
A.~J.~Buras and M.~Munz,
Phys.\ Rev.\ D {\bf 52}, 186 (1995)
[hep-ph/9501281].

\bibitem{pdg}
D.~E.~Groom {\it et al.}  [Particle Data Group Collaboration],
Eur.\ Phys.\ J.\ C {\bf 15}, 1 (2000).

\bibitem{ligeti}
Z.~Ligeti and M.~B.~Wise,
Phys.\ Rev.\ D {\bf 53}, 4937 (1996) [hep-ph/9512225].

\bibitem{thorsten}
M.~Beneke and T.~Feldmann, Nucl.\ Phys.\ B {\bf 592}, 3 (2001)
[hep-ph/0008255].


\end{thebibliography}
\end{document}